\documentclass{aa}

\usepackage{amsmath}
\usepackage{amssymb}
\usepackage{upgreek}
\usepackage[varg]{txfonts}
\usepackage{epsfig,graphicx}
\usepackage{xcolor}
\usepackage[colorlinks=true, linkcolor=blue, citecolor=blue, urlcolor=blue]{hyperref}
\usepackage{natbib}
\bibpunct{(}{)}{;}{a}{}{,} 
\renewcommand{\Re}{R_{\text{e}}}

\defcitealias{1991A&A...249...99C}{Paper~I}
\defcitealias{1997A&A...321..724C}{Paper~II}
\defcitealias{2019A&A...626A.110B}{Paper~III}

\begin{document}

\title{Stellar systems following the $R^{1/m}$ luminosity law}
\subtitle{IV. The total energy and the central concentration of galaxies}

\author{Maarten Baes\inst{\ref{UGent}} \and Luca Ciotti\inst{\ref{Bologna}}}

\institute{
Sterrenkundig Observatorium, Universiteit Gent, Krijgslaan 281 S9, 9000 Gent, Belgium
\label{UGent}
\and
Dipartimento di Fisica e Astronomia, Universit\`{a} di Bologna, Via Piero Gobetti 93/2, Bologna, Italy
\label{Bologna}
}

\abstract{We expand our previous analytical and numerical studies of the family of S\'ersic models, routinely used to describe early-type galaxies and the bulges of spiral galaxies. In particular, we focus on the total energy budget, an important dynamical property that has not been discussed in detail in previous works. We use two different methods to calculate the total energy for the S\'ersic model family, resulting in two independent expressions that can be used along the entire sequence of S\'ersic models. We use these expressions to investigate whether the Spitzer concentration index is a reliable measure for the intrinsic 3D concentration of galaxies, and we conclude that it is not a very useful measure for the central concentration. The popular Third Galaxy Concentration index, on the other hand, is shown to be a reliable measure for the intrinsic 3D concentration, even though it is based on the surface brightness distribution and not on the intrinsic 3D density.}


\maketitle

\section{Introduction}

Over the past decades, the S\'ersic model has become the preferred model to describe the surface brightness profiles of early-type galaxies and spiral galaxies bulges \citep[e.g.,][]{1988MNRAS.232..239D, 1993MNRAS.265.1013C, 2001A&A...368...16M, 2003AJ....125.2936G, 2006MNRAS.371....2A, 2009MNRAS.393.1531G, 2012MNRAS.427.1666B, 2012MNRAS.421.1007K, 2012ApJS..203...24V, 2015ApJS..219....4S, 2016MNRAS.462.1470L}. The model is the prime component of all modern galaxy profile fitting codes \citep{2002AJ....124..266P, 2010AJ....139.2097P, 2014AstBu..69...99M, 2015ApJ...799..226E, 2017MNRAS.466.1513R}. It is hence not surprising that the properties of the S\'ersic model have been examined in large detail in the past three decades. 

As the model is defined by means of the surface brightness profile, many of the projected, i.e., on-sky, properties can be expressed analytically (\citealt{1991A&A...249...99C}, hereafter \citetalias{1991A&A...249...99C}; \citealt{1999A&A...352..447C}; \citealt{2001MNRAS.326..869T}). A compendium of the most important photometric properties has been presented by \citet{2005PASA...22..118G}, and the gravitational lensing characteristics are discussed by \citet{2004A&A...415..839C} and \citet{2007JCAP...07..006E}.

An annoying aspect of the S\'ersic model is that the standard Abel deprojection of the surface brightness profile does not yield a closed expression for the density in terms of elementary functions or even in terms of standard special functions. Several authors proposed approximations for the spatial density of the S\'ersic model \citep{1997A&A...321..111P, 1999MNRAS.309..481L, 2002MNRAS.333..510T}. It turns out that closed expressions for the density and related properties can be derived using Mellin integral transforms. The resulting expressions are written in terms of the Fox $H$ function, or the Meijer $G$ function for integer and half-integer values of $m$ \citep{2002A&A...383..384M, 2011A&A...525A.136B, 2011A&A...534A..69B}.

The dynamical properties of the S\'ersic model were first investigated in the first two papers of this series (\citetalias{1991A&A...249...99C}; \citealt{1997A&A...321..724C}, hereafter \citetalias{1997A&A...321..724C}). These papers focused on relatively large S\'ersic indices ($m\geqslant2$). These studies were extended by \citet[][hereafter \citetalias{2019A&A...626A.110B}]{2019A&A...626A.110B}, where we considered the entire range of S\'ersic indices, and particularly focused on small values of $m$, appropriate for low-mass and dwarf ellipticals. An important result of these studies is that all S\'ersic models with $m\geqslant \tfrac12$ can be supported by an isotropic velocity dispersion tensor, and that these isotropic models are stable to both radial and non-radial perturbations. S\'ersic models with smaller values of $m$, however, cannot be supported by an isotropic velocity dispersion tensor.

A dynamical property of the S\'ersic models that has not been discussed analytically is their total energy. For example, the total energy budget of an equilibrium dynamical model is relevant for numerical studies, as it sets the preferred length scale for Monte Carlo or $N$-body simulations. The need for a consistent set of standard units for cluster simulations has been advocated since the 1970s, and the most popular system that has emerged is the system of so-called standard $N$-body units \citep{1971Ap&SS..14..151H, 1979ApJ...234.1036C, 1986LNP...267..233H}.\footnote{The use of this unit system has been strongly advocated by \citet{1986LNP...267..233H}, and as a result, these standard units have sometimes been called Heggie units. In 2014, Douglas Heggie proposed the name H\'enon units to commemorate the original proposer.} This unit system is defined by the requirements $G=M=1$, $E_{\text{tot}} = -\frac14$, or equivalently, uses the virial radius as the length unit.

Moreover, from the theoretical point of view, the total energy budget is one of the ingredients required to calculate  the concentration parameter introduced by \citet{1969ApJ...158L.139S}. Contrary to other concentration indices \citep[e.g.,][]{2001MNRAS.326..869T, 2001AJ....122.1707G, 2018MNRAS.477.2399A}, it is based on the intrinsic 3D density distribution, rather than on the light distribution on the plane of the sky. In the past few years, the interest in the central light (or mass) concentration of galaxies has only increased, thanks to a number of scaling relations between the central concentration and other galactic parameters, including velocity dispersion, supermassive black hole mass, optical-to-X-ray flux ratio and nuclear radio emission \citep[e.g.,][]{2001AJ....122.1707G, 2001ApJ...563L..11G, 2009ApJ...702L..51P, 2018MNRAS.477.2399A}. If physical processes in the evolution of a galaxy affect the mass/light concentration in a galaxy, one would primarily expect correlations that involve concentration indices based on the intrinsic density. It does hence make sense to investigate the relation between intrinsic and projected concentration indices for the S\'ersic model, in particular for the low $m$ regime where the intrinsic density distribution shows interesting characteristics \citepalias{2019A&A...626A.110B}.

The goal of our study is two-fold. Firstly, we want to extend the body of analytical studies on the S\'ersic model by providing a closed expression for the total energy. Secondly, we use these expressions to compare the intrinsic Spitzer concentration index to the commonly used TGC light concentration index \citep{2001MNRAS.326..869T}, to find out how we can best parameterise the intrinsic 3D concentration. In Section~{\ref{Sersic.sec}} we summarise some general properties of the family of S\'ersic models. In Section~{\ref{Etot.sec}} we compute the total energy of the family of S\'ersic models using two different approaches: the strip brightness approach and the Mellin integral transform framework. In Section~{\ref{CI.sec}} we use these results to compare 2D and 3D concentration indices for the S\'ersic model, and we compare the S\'ersic model with other popular families of spherical dynamical models. Our results are summarised in Section~{\ref{Summary.sec}}.

\section{The S\'ersic model}
\label{Sersic.sec}

The S\'ersic model is defined by the surface brightness profile
\begin{equation}
I(R) = I_0\exp\left[-b\left(\frac{R}{\Re}\right)^{1/m}\right].
\label{sersic}
\end{equation}
It is a three-parameter family with $I_0$ the central surface brightness, $\Re$ the effective radius, and $m$ the so-called S\'ersic index. The parameter $b=b(m)$ is not a free parameter in the model, but a dimensionless parameter that is set such that $\Re$ corresponds to the isophote that contains half of the emitted luminosity. For a given $m$, the corresponding value of $b$ can be found by solving a non-algebraic equation, and various interpolation formulae have been presented in the literature (\citealt{1989woga.conf..208C, 1997A&A...321..111P, 2003ApJ...582..689M}; \citetalias{1991A&A...249...99C, 2019A&A...626A.110B}). In particular, we recall the exact asymptotic formulae for large and small values of $m$ (\citealt{1999A&A...352..447C}, \citetalias{2019A&A...626A.110B}).

Instead of the central surface brightness $I_0$ we can also use the total luminosity $L$ as a free parameter. The connection between both quantities is
\begin{equation}
I_0 = \frac{b^{2m}}{2\pi\,m\,\Gamma(2m)}\,\frac{L}{\Re^2}. 
\label{I0}
\end{equation}
For more formulae related to the S\'ersic model, and for figures illustrating how the most important properties vary as a function of $m$, we refer to \citetalias{1991A&A...249...99C}, \citetalias{2019A&A...626A.110B}, and \citet{2005PASA...22..118G}. 

\section{The total energy of the S\'ersic model}
\label{Etot.sec}

For a spherically symmetric system characterised by a mass density $\rho(r)$ and a gravitational potential $\Phi(r)$, the expression for the total energy $E_{\text{tot}}$ is given by 
\begin{equation}
E_{\text{tot}} = \tfrac12\,U_{\text{tot}}
= 
\pi\int_0^\infty \rho(r)\,\Phi(r)\,r^2\,{\text{d}}r.
\label{Etot-def}
\end{equation}
In this expression, $U_{\text{tot}}$ represents the total potential energy of the system, and the equality $E_{\text{tot}} = \tfrac12\,U_{\text{tot}}$ is a manifestation of the virial theorem \citep[e.g.,][]{2008gady.book.....B}. An alternative expression for $E_{\text{tot}}$ is based on the cumulative mass density $M(r)$ instead of the gravitational potential,
\begin{equation}
E_{\text{tot}} 
= 
-2\pi\,G\int_0^\infty \rho(r)\,M(r)\,r\,{\text{d}}r.
\label{Etot-def2}
\end{equation}
Considering that the spatial density of the S\'ersic model, obtained from an Abel inversion of Eqn.~(\ref{sersic}), is not an elementary function, (and so even less the derived quantities such as the potential and the cumulative mass), and that the two integrals above involve products of those functions, it seems natural that the only approach to their evaluation is numerical integration. Starting from the surface brightness profile, expressions (\ref{Etot-def}) and (\ref{Etot-def2}) are five-dimensional and four-dimensional integrals, respectively. Quite surprisingly, in the following we show that in fact it is possible to obtain two different expressions for the total energy, by using the strip brightness quantity introduced by \citet{1954AJ.....59..273S} and by direct integration using advanced special functions.

\subsection{Calculation using the strip brightness}

A first method to calculate the total energy uses the strip brightness ${\mathcal{S}}(z)$, a quantity defined so that ${\mathcal{S}}(z)\,{\text{d}}z$ is the total luminosity in a strip of width ${\text{d}}z$ on the plane of the sky that passes a distance $z$ from the centre of the system. For a spherically symmetric system, the strip brightness can be written as \citep{1954AJ.....59..273S}
\begin{equation}
{\mathcal{S}}(z) = 2\pi \int_z^\infty \nu(r)\,r\,{\text{d}}r
\end{equation}
where $\nu(r)$ is the luminosity density. An equivalent expression for ${\mathcal{S}}(z)$ is
\begin{equation}
{\mathcal{S}}(z) = 2\int_z^\infty \frac{I(R)\,R\,{\text{d}}R}{\sqrt{R^2-z^2}}.
\label{strip}
\end{equation}
The equivalence of these two expression can easily be demonstrated by inserting the projection equation
\begin{equation}
I(R) = 2\int_R^\infty \frac{\nu(r)\,r\,{\text{d}}r}{\sqrt{R^2-r^2}}
\end{equation}
into expression~(\ref{strip}) and changing the order of the resulting double integral \citep[see also][Problem 1.3]{2008gady.book.....B}.

\citet{1954AJ.....59..273S} demonstrated that $E_{\text{tot}}$ can be calculated from the strip brightness using 
\begin{equation}
E_{\text{tot}} = -G\,\Upsilon^2 \int_0^\infty {\mathcal{S}}^2(z)\,{\text{d}}z.
\label{Etotstrip}
\end{equation}
with $\Upsilon$ the mass-to-light ratio of the system. We now elaborate on the previous identity, following a path apparently unnoticed in \citet{1954AJ.....59..273S}. We will first obtain a generic two-dimensional integral expression for $E_{\text{tot}}$ in terms of the surface brightens profile, and then in the special case of the S\'ersic profile we will show that the integral can be in fact be reduced to a one-dimensional integral. We proceed as follows. Inserting Eqn.~(\ref{strip}) into (\ref{Etotstrip}), one finds a triple integral
\begin{equation}
E_{\text{tot}} = -4\,G\,\Upsilon^2 
\int_0^\infty {\text{d}}z
\int_z^\infty \frac{I(x)\,x\,{\text{d}}x}{\sqrt{x^2-z^2}}
\int_z^\infty \frac{I(y)\,y\,{\text{d}}y}{\sqrt{y^2-z^2}}.
\end{equation}
Changing the order of integrations, one finds after some calculation
\begin{align}
E_{\text{tot}} &= -8\,G\,\Upsilon^2 \int_0^\infty I(x)\,{\text{d}}x
\int_0^x I(y)\,{\mathbb{K}}\left(\frac{y}{x}\right) {\text{d}}y,
\nonumber \\
&= -8\,G\,\Upsilon^2 \int_0^{\pi/4} {\mathbb{K}}(\tan\phi) f(\phi) \sin\phi\,{\text{d}}\phi,
\label{Etotf}
\end{align}
with ${\mathbb{K}}(k)$ the complete elliptic integral of the first kind, and with the definition
\begin{equation}
f(\phi) = \int_0^\infty I(R\cos\phi)\,I(R\sin\phi)\,R^2\,{\text{d}}R.
\label{poi}
\end{equation}
Up to now, we have used generic formulae, and not yet used the specific form of the S\'ersic surface brightness profile. This expression shows that, by using the strip brightness function introduced by \citet{1954AJ.....59..273S} and repeated exchanges in the integration order, the total energy of any generic spherical model defined by a surface brightness density $I(R)$ can be always reduced to a two-dimensional integral. Interestingly, for the S\'ersic model, one of the two integrals can be evaluated analytically. Indeed, with expression~(\ref{sersic}), $f(\phi)$ becomes
\begin{equation}
f(\phi) = I_0^2 \int_0^\infty \exp\left[-b\left(\frac{R}{\Re}\right)^{1/m}\Omega\right] R^2\,{\text{d}}R
= \frac{I_0^2\,\Re^3 m\,\Gamma(3m)}{b^{3m}\,\Omega^{3m}}.
\end{equation}
with the quantity $\Omega$ defined as $\Omega = \cos^{1/m}\phi+\sin^{1/m}\phi$. Inserting this expression into Eqn.~(\ref{Etotf}) reduces the expression for the total energy to a relatively single expression with that involves just a single integration. Setting $k=\tan\phi$, and using expression (\ref{I0}), we obtain
\begin{equation}
E_{\text{tot}} = -\frac{2\,\Gamma(3m)\,b^m}{\pi^2\,m\,\Gamma^2(2m)}\,\frac{G M^2}{\Re} 
\int_0^1 \frac{{\mathbb{K}}(k)\,k\,{\text{d}}k}{(1+k^{1/m})^{3m}}.
\label{Etot-K}
\end{equation}
with $M = \Upsilon\,L$ the total mass of the system. Many different integrals of the complete elliptic integral of the first kind can be evaluated exactly \citep[e.g.,][]{Glasser1976, Cvijovic1999, 2007tisp.book.....G}. Unfortunately, the integral in expression (\ref{Etot-K}) is not found among these lists. It is easily evaluated numerically, however, as the integrand is well-behaved over the entire integration domain.

\begin{table*}
\caption{Numerical values for the total energy $E_{\text{tot}}$, the gravitational radius $r_{\text{G}}$, the half-mass radius $r_{\text{h}}$, the Spitzer concentration index $C_{\text{S}}$, the TGC concentration index, and the 3D version of the TGC (${\text{TGC}}_{\text{3D}}$), as a function of $m$.}
\label{Etot.tab}
\centering
\begin{tabular}{ccccccc}
  \hline \hline \\
  $m$ & $\dfrac{E_{\text{tot}}}{GM^2\!/\Re}$ &  $\dfrac{r_{\text{G}}}{\Re}$
  & $\dfrac{r_{\text{h}}}{\Re}$ & $C_{\text{S}}$ & TGC & ${\text{TGC}}_{\text{3D}}$ \\ \\ \hline \\
$0.0$ & $-0.19105306$ & $2.6170741$ & $1.2936816$ & $0.49432364$ & $0.11111111$ & $0.024772219$ \\
$0.5$ & $-0.16607062$ & $3.0107674$ & $1.3064032$ & $0.43391038$ & $0.14825058$ & $0.066299036$ \\
$1.0$ & $-0.15734503$ & $3.1777299$ & $1.3248257$ & $0.41690949$ & $0.21747529$ & $0.15312211$ \\
$1.5$ & $-0.15452645$ & $3.2356921$ & $1.3337332$ & $0.41219411$ & $0.27914764$ & $0.22956694$ \\
$2.0$ & $-0.15453561$ & $3.2355003$ & $1.3389685$ & $0.41383663$ & $0.33033179$ & $0.29121676$ \\
$2.5$ & $-0.15628600$ & $3.1992630$ & $1.3424141$ & $0.41960105$ & $0.37280793$ & $0.34111712$ \\
$3.0$ & $-0.15928070$ & $3.1391122$ & $1.3448539$ & $0.42841854$ & $0.40850538$ & $0.38222620$ \\
$3.5$ & $-0.16325932$ & $3.0626122$ & $1.3466723$ & $0.43971363$ & $0.43893749$ & $0.41671952$ \\
$4.0$ & $-0.16807502$ & $2.9748621$ & $1.3480801$ & $0.45315716$ & $0.46522882$ & $0.44613983$ \\
$4.5$ & $-0.17364254$ & $2.8794787$ & $1.3492021$ & $0.46855775$ & $0.48821266$ & $0.47158994$ \\
$5.0$ & $-0.17991325$ & $2.7791172$ & $1.3501174$ & $0.48580801$ & $0.50851353$ & $0.49387312$ \\
$6.0$ & $-0.19447936$ & $2.5709669$ & $1.3515209$ & $0.52568584$ & $0.54285984$ & $0.53118369$ \\
$7.0$ & $-0.21173840$ & $2.3614045$ & $1.3525465$ & $0.57277206$ & $0.57094111$ & $0.56135261$ \\
$8.0$ & $-0.23180415$ & $2.1569933$ & $1.3533288$ & $0.62741447$ & $0.59444205$ & $0.58638759$ \\
$9.0$ & $-0.25488801$ & $1.9616458$ & $1.3539452$ & $0.69020878$ & $0.61447913$ & $0.60759027$ \\
$10.0$ & $-0.28127786$ & $1.7776017$ & $1.3544433$ & $0.76194984$ & $0.63182362$ & $0.62584461$ \\ \\
\hline \hline \\
\end{tabular}
\end{table*}

\subsection{Calculation using advanced special functions}

A second method to calculate $E_{\text{tot}}$ for the S\'ersic model is by using the analytical expressions for the density and related properties derived by \citet{2011A&A...525A.136B} and \citet{2011A&A...534A..69B} in terms of the Fox $H$ function. The general expression for the density is \citepalias{2019A&A...626A.110B}
\begin{equation}
\rho(r) =
\frac{b^{3m}}{\pi^{3/2}\,\Gamma(2m)}\,\frac{M}{\Re^3}\,u^{-1}\,
H^{2,0}_{1,2} \left[ \left.\begin{matrix} (0,1) \\ (0,2m), (\tfrac12,1) \end{matrix} \,\right| u^2 \right],
\label{rho-FoxH}
\end{equation}
where we have used the dimensionless spherical radius
\begin{equation}
u=\frac{b^m r}{\Re}
\end{equation}
The corresponding mass profile is
\begin{equation}
M(r) 
=
\frac{2M}{\sqrt{\pi}\, \Gamma(2m)}\,u^2\,
H^{2,1}_{2,3} \left[ \left.\begin{matrix} (0,1), (0,1) \\ (0,2m), (\tfrac12,1), (-1,1) \end{matrix}\,\right| u^2 \right].
\label{M-FoxH}
\end{equation}
When we substitute the expressions (\ref{rho-FoxH}) and (\ref{M-FoxH}) in the definition (\ref{Etot-def}), we find the total energy
\begin{multline}
E_{\text{tot}} = -\frac{b^m}{2\pi\, \Gamma^2(2m)}\,\frac{GM^2}{\Re}
\\ \qquad\quad\times  
\int_0^\infty  H^{2,1}_{2,3}\left[\left.\begin{matrix} (0,1), (0,1) \\ (0,2m), (-\tfrac12,1), (-1,1) \end{matrix}\,\right| z \right]
\\ \times 
H^{2,0}_{1,2}\left[\left.\begin{matrix} (0,1) \\ (0,2m), (\tfrac12,1) \end{matrix}\,\right| z \right] z^{1/2}\, {\text{d}}z.
\end{multline}
This integral can be evaluated using the standard integration formula for a product of two Fox $H$ functions \citep{2009hfta.book.....M}, and after some simplifications one obtains
\begin{multline}
E_{\text{tot}} = -\frac{b^m}{2\pi\, \Gamma^2(2m)}\,\frac{GM^2}{\Re}
\\ \times
H^{2,2}_{3,3} \left[\left.\begin{matrix} (1-3m,2m),(0,1),(0,1) \\ (0,2m), (-\tfrac12,1), (-\tfrac12,1) \end{matrix} \,\right| 1 \right].
\label{Etot-FoxH}
\end{multline}
For integer and half-integer values of $m$, the Fox $H$ function in expression (\ref{Etot-FoxH}) can be reduced to Meijer $G$ functions. This reduction is based on the integral representations of the Meijer $G$ and Fox $H$ function, and Gauss' multiplication theorem. The result reads
\begin{multline}
E_{\text{tot}} = -\frac{(2m)^{3m-1}\,b^m}{(2\pi)^{2m}\,\Gamma^2(2m)}\,\frac{GM^2}{\Re}\,
\\ \times 
G^{2m,2m}_{2m,2m} \!\left[\left. \begin{matrix} -\tfrac{m+1}{2m},-\tfrac{m+2}{2m},\ldots,-\tfrac{3m-1}{2m},0 \\ \tfrac{1}{2m},\tfrac{2}{2m}\ldots,\tfrac{2m-1}{2m},-\tfrac12 \end{matrix} \,\right|1 \right].
\label{Etot-MeijerG}
\end{multline}

\subsection{Numerical values}

In Table~{\ref{Etot.tab}} we tabulate the value of the total energy and the gravitational radius $r_{\text{G}}$, defined through the relation
\begin{equation}
2E_{\text{tot}} = U_{\text{tot}} = -\frac{GM^2}{r_{\text{G}}},
\label{rG}
\end{equation}
for a number of values between $m=0$ and $m=10$. These values have been calculated through both expressions~(\ref{Etot-K}) and (\ref{Etot-MeijerG}) with 15 significant digits, and are found to be in perfect agreement. We also find perfect agreement with the analytical results for the few special cases for which $E_{\text{tot}}$ can be calculated analytically, i.e., for $m=0$, $\tfrac12$ and $1$ (Appendix~{\ref{SpecialCases.sec}}). We can hence conclude that both the expressions are equivalent, or that the integral in Eqn.~(\ref{Etot-K}) can be evaluated exactly as 
\begin{multline}
\int_0^1 \frac{{\mathbb{K}}(k)\,k\,{\text{d}}k}{(1+k^{1/m})^{3m}}
\\=
\frac{\pi\,m}{4\,\Gamma(3m)}
H^{2,2}_{3,3} \left[\left.\begin{matrix} (1-3m,2m),(0,1),(0,1) \\ (0,2m), (-\tfrac12,1), (-\tfrac12,1) \end{matrix} \,\right| 1 \right].
\end{multline}
The values for $E_{\text{tot}}$ for $2\leqslant m\leqslant10$ are in good agreement to those listed in \citetalias{1991A&A...249...99C}, obtained through numerical integration.

\section{Discussion}
\label{CI.sec}

\subsection{Central concentration of the S\'ersic models}

The calculation of the total energy of the family of S\'ersic models is primarily important in the discussion on the central concentration of galaxies. The degree to which light or mass is centrally concentrated is an important diagnostic for galaxies. The importance is obvious when one considers the many physical galaxy properties that correlate with (different measures of) the galaxy light concentration, including total luminosity \citep{1993MNRAS.265.1013C, 2001AJ....122.1707G}, velocity dispersion \citep{2001AJ....122.1707G}, Mg/Fe abundance ratio \citep{2004ApJ...601L..33V}, central supermassive black hole mass \citep{2001ApJ...563L..11G, 2018MNRAS.477.2399A}, cluster local density \citep{2002ApJ...573L...9T}, and emission at radio and X-ray wavelengths \citep{2009ApJ...702L..51P, 2018MNRAS.477.2399A}. This has inspired several teams to propose galaxy concentration as an important parameter in automated galaxy classification schemes \citep{1993MNRAS.264..832D, 1994ApJ...432...75A, 2000AJ....119.2645B, 2003ApJS..147....1C}.

There are many different ways in which the central concentration of galaxies can be estimated or parameterised. A number of concentration indices, such as the widely used $C_{31}$ index, are defined as the ratio of radii that contain certain fractions of the total galaxy luminosity \citep{1977egsp.conf...43D, 1985ApJS...59..115K, 2000AJ....119.2645B}. Other concentration indices are based on the ratio of the luminous flux enclosed by two different apertures \citep{1984ApJ...280....7O, 1993MNRAS.264..832D}. Possibly the most commonly used measure for the central light concentration of galaxies today is the Third Galaxy Concentration index or TGC index, introduced by \citet{2001MNRAS.326..869T} as the ratio between the flux within the isophote at a radius $\alpha\Re$ -- with $\alpha$ a number between 0 and 1 -- and the flux within the isophote at $\Re$, 
\begin{equation}
\label{TGCdef}
{\text{TGC}} = \frac{S(\alpha\Re)}{S(\Re)}
\qquad{\text{with}}\quad
S(R) = 2\pi\int_0^R I(R')\,R'\,{\text{d}}R'.
\end{equation}
For the family of S\'ersic models, the TGC index can be calculated analytically \citep{2001MNRAS.326..869T, 2005PASA...22..118G},
\begin{equation}
{\text{TGC}} = \frac{\gamma(2m,b\alpha^{1/m})}{\gamma(2m,b)}.
\label{TGCsersic}
\end{equation}
where $\gamma(s,x)$ is the lower incomplete gamma function. Note that expression (\ref{TGCsersic}) only depends on $\alpha$ and the S\'ersic index $m$; there is no dependency on effective radius, luminosity, or central surface brightness. In the remainder of this paper, we will always use $\alpha=\tfrac13$, the value generally adopted \citep[e.g.,][]{2001MNRAS.326..869T, 2001ApJ...563L..11G, 2010A&A...512A..35P}. We have, however, repeated the entire analysis for different values of $\alpha$, and have found that our results and conclusions are not sensitive to the particular choice of $\alpha$.

\begin{figure}
\centering
\includegraphics[width=0.85\columnwidth]{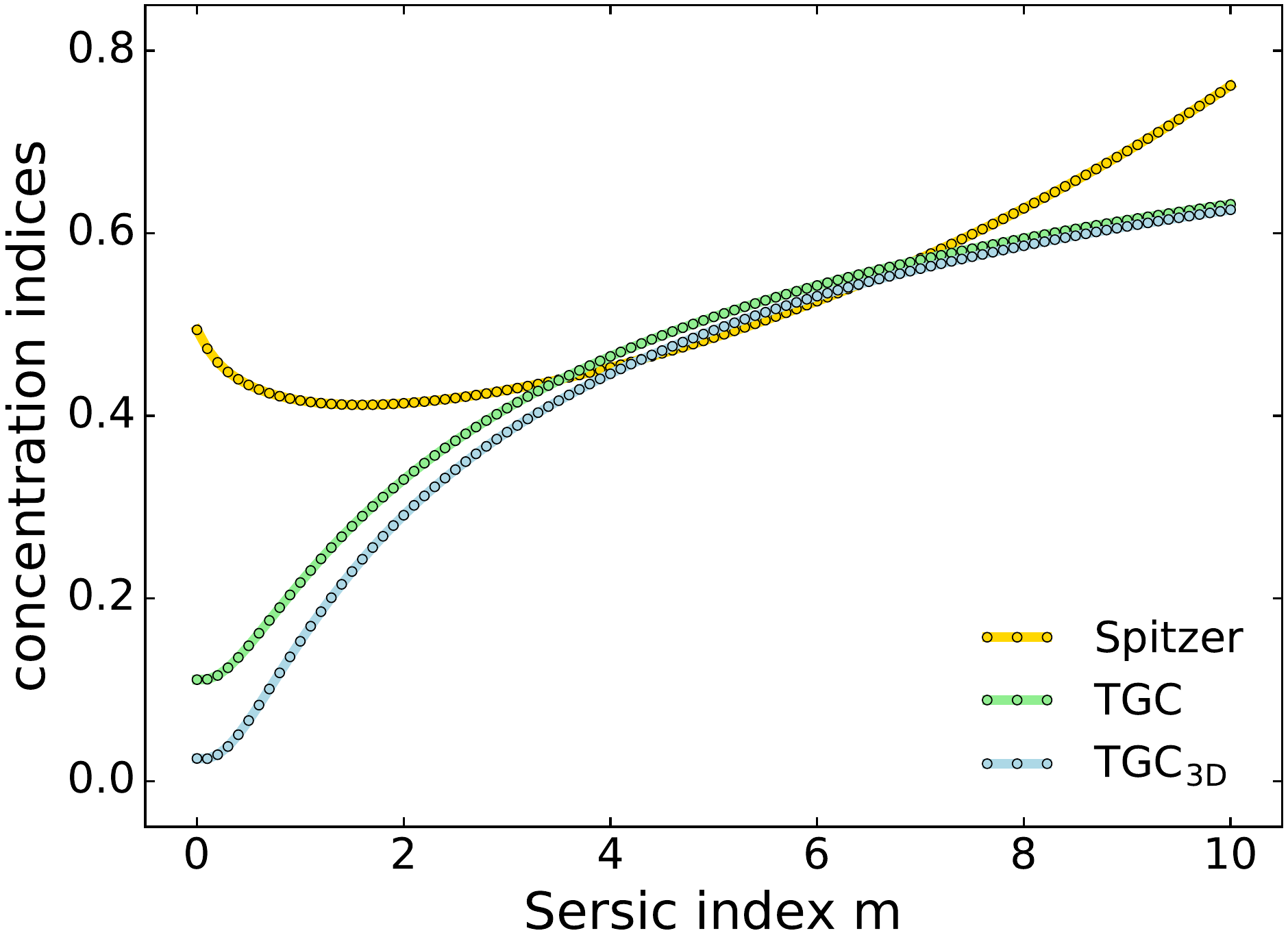}\hspace*{2em}
\caption{The dependence of the Spitzer concentration index (\ref{TGCdef}), the TGC index (\ref{TGCdef}) and the 3D TGC index (\ref{TGC3Ddef}) on the S\'ersic index $m$.}  
\label{Sersic-Concentration.fig}
\end{figure}

As already shown by \citet{2001MNRAS.326..869T} and \citet{2001AJ....122.1707G}, the TGC index is a monotonically increasing function of $m$. In the limit of $m\rightarrow0$, the surface brightness profile is a uniform disc on the sky \citepalias{2019A&A...626A.110B}, and it is easy to see that ${\text{TGC}} \rightarrow \tfrac19$. In Fig.~{\ref{Sersic-Concentration.fig}}, the green line shows the TGC index as a function of the S\'ersic index $m$.

\begin{figure*}
%\centering
\includegraphics[width=0.95\textwidth]{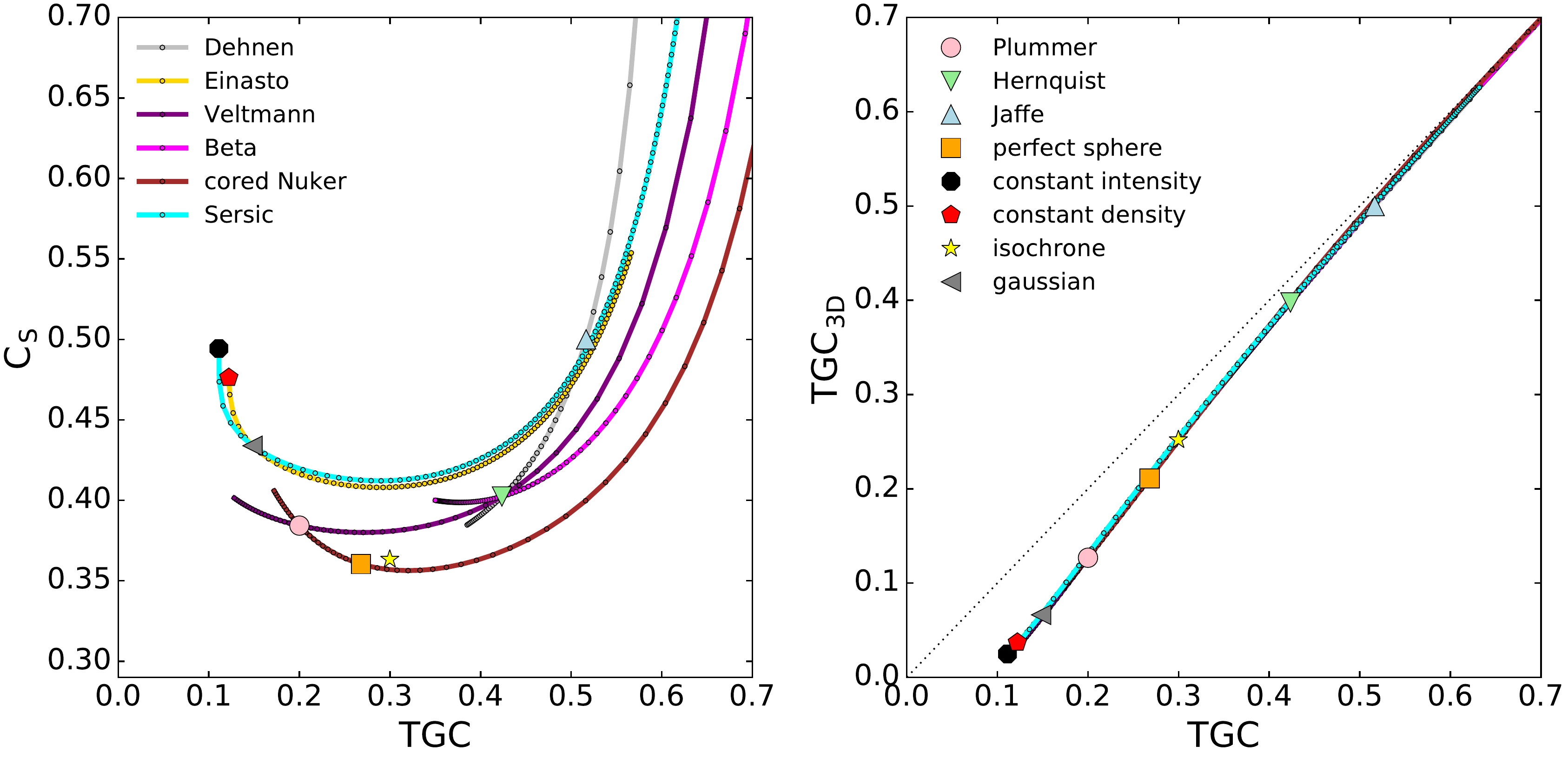}
\caption{The location of the S\'ersic model family, as well as most other popular spherical toy models, in the $({\text{TGC}},C_{\text{S}})$ and the $({\text{TGC}},{\text{TGC}}_{\text{3D}})$ planes. The models form banana-shaped trails in the former plane, whereas they are located on an almost perfect one-to-one relation in the latter plane. This shows that the Spitzer concentration index is a poor indicator of the intrinsic 3D concentration, whereas the TGC is a very accurate one.}  
\label{ConcentrationIndices.fig}
\end{figure*}

A potential caveat of concentration indices as the $C_{31}$ or TGC indices is that they are based on the observed, projected distribution of light on the plane of the sky. A physical  characterisation of the central concentration of light (or mass) should in principle be based on the intrinsic 3D density distribution. A characterisation that satisfies this requirement is the Spitzer concentration index \citep{1969ApJ...158L.139S}, defined almost half a centuary ago as the ratio between the half-mass radius and the gravitational radius,
\begin{equation}
C_{\text{S}} = \frac{r_{\text{h}}}{r_{\text{G}}}.
\label{CSdef}
\end{equation}
The half-mass radius $r_{\text{h}}$ is obviously the radius of the spherical volume that contains half of the total mass, and the gravitational radius is defined through Eqn.~(\ref{rG}).

Table~2 of \citetalias{1991A&A...249...99C} lists numerical approximations for the Spitzer concentration index for a number of S\'ersic models with $m\geqslant2$. It was noted that $C_{\text{S}}$ is an monotonically increasing function of $m$, as one would expect. This behaviour does not extend over the entire range of S\'ersic indices, however. The yellow line in Fig.~{\ref{Sersic-Concentration.fig}} shows how $C_{\text{S}}$ varies with $m$ between 0 and 10, and numerical values are listed in the fifth column of Table~{\ref{Etot.tab}}. Contrary to the TGC index, Spitzer concentration index is not a monotonically increasing function of $m$. For $m\gtrsim1.6$ it does increase with increasing $m$, in agreement with the observation in \citetalias{1991A&A...249...99C}. For values of $m\lesssim1.6$, $C_{\text{S}}$ increases again for decreasing $m$ with a rate that is quite steep due to the strong variation of the total energy. At $m=0$, a limiting value $C_{\text{S}} \approx 0.4943$ is reached, which would imply that the constant intensity model would be more centrally concentrated than a de Vaucouleurs model.

A logical consequence is that the TGC and Spitzer concentration indices are not correlated for the family of S\'ersic models. In the left panel of Fig.~{\ref{ConcentrationIndices.fig}} we show the position of the family of S\'ersic models in the plane formed by these two indices. The sequence of models forms a banana-shaped trail in this diagram. This diagram suggests that the Spitzer index is a poor metric to indicate the central mass/light concentration in galaxies.

\subsection{Comparison to other models}

In order to further investigate the usefulness of the Spitzer index as an indicator for the central concentration, we have also calculated the TGC and $C_{\text{S}}$ indices for a number of other popular families of toy models that are often used to represent galaxies. These families are also shown in the left panel of Fig.~{\ref{ConcentrationIndices.fig}}. Apart from the sequence that corresponds to the S\'ersic models, this plot also contains the $\gamma$- or Dehnen models \citep{1993MNRAS.265..250D, 1994AJ....107..634T}, the $\beta$-models \citep{1996MNRAS.278..488Z}, the Veltmann or hypervirial models \citep{1979AZh....56..976V, 2005MNRAS.360..492E}, the Einasto models \citep{1965TrAlm...5...87E, 2005MNRAS.358.1325C}, and the cored Nuker or Zhao $(\tfrac12,\beta,0$)-models \citep{1996MNRAS.278..488Z}. Some well-known specific models are also indicated: the Plummer model \citep{1911MNRAS..71..460P, 1987MNRAS.224...13D}, H\'enon's isochrone sphere \citep{1959AnAp...22..126H}, the Hernquist model \citep{1990ApJ...356..359H}, the Jaffe model \citep{1983MNRAS.202..995J}, the perfect sphere \citep{1985MNRAS.216..273D}, the gaussian model (Section~{\ref{GaussianModel.sec}}), the constant density sphere \citep{2008gady.book.....B}, and the constant intensity sphere \citepalias{2019A&A...626A.110B}. Most of these models belong to one or more of the families mentioned above. In particular, the Hernquist model lies at the intersection of the Veltmann, $\beta$- and Dehnen models, the Plummer sphere belongs to the Veltmann and cored Nuker families, and the gaussian model is common between the S\'ersic and Einasto sequences. For all of these models, the total energy budget can be calculated analytically (Appendix~{\ref{AllModels.sec}}).

It is quite interesting to see that all of these different models occupy a relatively narrow region in the $({\text{TGC}},C_{\text{S}})$ plane.  This is remarkable, given the large variety in central density slopes between these models, ranging from models with a constant central density to models with a strong density cusp. Furthermore, it is clear that the banana-shaped trail of the S\'ersic models is not unique to this specific family of models. On the contrary, it seems to be a general feature: the Veltmann, Einasto, $\beta$ and cored Nuker models show the same behaviour. Among the models with the lowest $C_{\text{S}}$ values are the perfect sphere and H\'enon's isochrone sphere, two models with a central density core and a relatively shallow $r^{-4}$ fall-off at large radii. It does not make sense that these models, according to this concentration index, would be characterised as less centrally concentrated than the constant intensity sphere, in which the density actually increasing with increasing radius \citepalias{2019A&A...626A.110B}. In conclusion, the Spitzer index is not a very useful measure for the central concentration of dynamical models.

\subsection{The intrinsic 3D concentration}

If the Spitzer concentration index is not a useful measure for the intrinsic 3D concentration, which index should one use? Based on the monotonic dependence of the TGC index on $m$ for the family of S\'ersic models, one could imagine that the TGC index, while defined to measure the concentration of the surface brightness distribution on the sky, is also a suitable measure for the intrinsic 3D concentration. Similarly, for the family of Dehnen models, the TGC index increases monotonically with the central slope $\gamma$, which is a natural measure for the central concentration for this family. 

To test whether the TGC index is a reliable measure for the intrinsic 3D density concentration, we define a general 3D version of the TGC index as ratio between the mass contained within a sphere with radius $\alpha r_{\text{h}}$ and the mass contained within the half-mass radius $r_{\text{h}}$,
\begin{equation}
\label{TGC3Ddef}
{\text{TGC}}_{\text{3D}} = \frac{M(\alpha r_{\text{h}})}{M(r_{\text{h}})}
\qquad{\text{with}}\quad
M(r) = 4\pi\int_0^r \rho(r')\,r'^2\,{\text{d}}r'.
\end{equation}
Again we assume $\alpha=\tfrac13$. The blue line in Fig.~{\ref{Sersic-Concentration.fig}} shows that the ${\text{TGC}}_{\text{3D}}$ index varies monotonically as function of $m$, in a way that is very similar to the TGC index. 

The right panel of Fig.~{\ref{ConcentrationIndices.fig}} shows the correlation between the TGC and ${\text{TGC}}_{\text{3D}}$ indices, not only for the S\'ersic family, but for all the models also shown in the left panel. There is an almost perfect one-to-one correlation between both indices, over the different classes of models. For models with a small central concentration, such as the S\'ersic models with small $m$, the ${\text{TGC}}_{\text{3D}}$ index is systematically lower than the TGC index. As the models are more and more centrally concentrated, the difference between the two indices becomes smaller, and for very centrally concentrated systems, both indices converge to one. The bottomline is that the TGC index is a reliable measure for the intrinsic 3D concentration, and no separate index as the ${\text{TGC}}_{\text{3D}}$ index needs to be invoked to distinguish between 2D and 3D concentration of galaxies. It hence makes perfect sense to use the TGC index in statistical studies between global galaxy parameters.

\section{Summary}
\label{Summary.sec}

We have expanded our previous analytical and numerical studies of the family of S\'ersic models, and concentrated on the total energy budget. The main results of this dedicated study are the following.

Firstly, we explored the \citet{1954AJ.....59..273S} formalism of the strip brightness to calculate the total energy budget for the S\'ersic family. This results in a relatively simple expression that involves just a single integration. In a completely independent way, we obtained a closed expression for the total energy in terms of the Fox $H$ function, thanks to the closed expressions for density and related properties derived in our previous work \citep{2011A&A...525A.136B, 2011A&A...534A..69B}. In turn, this means that we have a closed form solution for the one-dimensional integral obtained along the previous approach. We were not able to find this expression in all the standard tables of special functions, and also the well known computer algebra systems were unable to compute the resulting integral. The two expressions are shown to be in agreement by performing numerical integration. We present a table with values for the total energy budget covering the entire range of S\'ersic parameters.

Subsequently, we use our calculations to investigate whether the Spitzer concentration index \citep{1969ApJ...158L.139S} is a reliable measure for the intrinsic 3D concentration of galaxies. We find that this is not the case: the index does not correlate with the S\'ersic parameter in the small $m$ range. More generally, we compare the Spitzer concentration index to the popular TGC index \citep{2001MNRAS.326..869T} for a wide range of spherical galaxy models, and find that these two indices do not correlate over the entire possible parameter space. We conclude that the Spitzer concentration index is not a very useful measure for the central concentration of dynamical models. On the other hand, we define a 3D version of the TGC index, and find an almost perfect correlation between the 2D and 3D versions, over a wide range of dynamical models. This implies that the TGC index is a reliable measure for the intrinsic 3D concentration, even though it is based on the surface brightness distribution and not on the intrinsic 3D density.

While this study is primarily a theoretical study, it also has a practical use for numerical studies of equilibrium dynamical models, as the total energy sets the preferred length scale in the standard or H\'enon unit system \citep{1971Ap&SS..14..151H, 1986LNP...267..233H}.

\bibliographystyle{aa}
\bibliography{SersicU}

\appendix
\section{Special cases}
\label{SpecialCases.sec}

There are a number of special S\'ersic models for which the total energy can be calculated analytically using elementary and/or simple special functions. 

\subsection{The exponential model $m=1$}
\label{ExponentialModel.sec}

For the special case $m=1$, the S\'ersic model has a simple exponential surface brightness profile, 
\begin{equation}
I(R) = I_0\,{\text{e}}^{-bR/\Re}.
\label{rho-exp}
\end{equation}
The spatial mass density $\rho(r)$ corresponding to this surface brightness profile, with $r$ the spherical radius, can be written in terms as
\begin{equation}
\rho(r) = \frac{b^3}{2\pi^2}\,\frac{M}{\Re^3}\,K_0(u),
\label{rho-exp}
\end{equation}
with $u=b\,r/\Re$, and $K_n(x)$ the modified Bessel function of the second kind of order $n$. After some calculation, one finds that the corresponding potential can be written as
\begin{gather}
\Phi(r) = -\frac{GMb}{\Re}
\left[ \left(\frac{2}{3\pi}\,u + L_2(u)\right)K_1(u) + L_1(u)\,K_2(u) \right],
\label{pot-exp}
\end{gather}
with $L_n(x)$ the modified Struve function of order $n$. Implausible as it may seem, if we substitute the density~(\ref{rho-exp}) and the potential~(\ref{pot-exp}) into Eqn.~(\ref{Etot-def}), the
resulting integral can be evaluated exactly as
\begin{equation}
E_{\text{tot}} = -\frac{3b}{32}\,\frac{GM^2}{\Re}.
\label{Etot-exp}
\end{equation}
On the other hand, if we set $m=1$ in the formula (\ref{Etot-MeijerG}), we get 
\begin{equation}
E_{\text{tot}} = -\frac{b}{\pi^2}\,\frac{GM^2}{\Re}\,
G^{2,2}_{2,2}\!\left[\left.\begin{matrix} -1,0 \\ \frac12,-\tfrac12 \end{matrix}\,\right|1\right].
\label{EtotG2222}
\end{equation}
All Meijer $G$ functions of the form $G^{2,2}_{2,2}$ can be written in terms of hypergeometric functions, and in this specific case one finds
\begin{equation}
G^{2,2}_{2,2}\!\left[\left.\begin{matrix} -1,0 \\ \frac12,-\tfrac12 \end{matrix}\,\right|z\right]
=
\frac{3\pi^2}{32}\,\frac{1}{\sqrt{z}}\,
{}_2F_1\left(\frac12,\frac32;3;1-z\right).
\end{equation}
Combining this with expression (\ref{EtotG2222}), we recover the simple result (\ref{Etot-exp}). Finally, if we set $m=1$ in expression (\ref{Etot-K}), we obtain the expression
\begin{equation}
E_{\text{tot}} = -\frac{4b}{\pi^2}\,\frac{GM^2}{\Re}
\int_0^1 \frac{{\mathbb{K}}(k)\,k\,{\text{d}}k}{(1+k)^3}.
\end{equation}
Unfortunately, neither Maple nor Mathematica manage to evaluate this integral symbolically, nor could this integral be evaluated using available lists of definite integrals involving the complete elliptic integral \citep{Glasser1976, Cvijovic1999}. It is, obviously, easy to check this result numerically.

\subsection{The gaussian model $m=\tfrac12$}
\label{GaussianModel.sec}

The S\'ersic model corresponding to $m=\tfrac12$ has a gaussian surface brightness profile, 
\begin{equation}
I(R) = I_0\,{\text{e}}^{-b\,R^2/\Re^2}.
\label{Igauss}
\end{equation}
Applying the standard deprojection formula, one finds also a gaussian density distribution,
\begin{equation}
\rho(r) = \frac{b^{3/2}}{\pi^{3/2}}\,\frac{M}{\Re^3}\,{\text{e}}^{-b\,r^2/\Re^2}.
\label{rho-gauss}
\end{equation}
If we set $u=\sqrt{b}\,r/\Re$, the potential can be written as
\begin{gather}
\Phi(r) = \frac{GM\!\sqrt{b}}{\Re}\,\frac{{\text{erf}}\,u}{u}.
\label{pot-gauss}
\end{gather}
If we substitute the expressions (\ref{rho-gauss}) and (\ref{pot-gauss}) in the definition~(\ref{Etot-def}), we obtain
\begin{equation}
E_{\text{tot}} = -\frac{\sqrt{b}}{2\!\sqrt{2\pi}}\,\frac{GM^2}{\Re}.
\label{Etot-gauss}
\end{equation}
On the other hand, if we set $m=\tfrac12$ in the general formula~(\ref{Etot-MeijerG}), we get
\begin{equation}
E_{\text{tot}} = -\frac{\sqrt{b}}{2\pi}\,\frac{GM^2}{\Re}\,
G^{1,1}_{1,1}\!\left[\left.\begin{matrix} 0 \\ -\frac12 \end{matrix}\,\right|1\right].
\end{equation}
Since 
\begin{equation}
G^{1,1}_{1,1}\!\left[\left.\begin{matrix} 0 \\ -\frac12 \end{matrix}\,\right|z\right]
=\sqrt{\frac{\pi}{z\,(1+z)}},
\end{equation}
we recover the same result (\ref{Etot-gauss}). Finally, substituting $m=\tfrac12$ into expression (\ref{Etot-K}) yields
\begin{equation}
E_{\text{tot}} = -\frac{2\sqrt{b}}{\pi^{3/2}}\,\frac{GM^2}{\Re}
\int_0^1 \frac{{\mathbb{K}}(k)\,k\,{\text{d}}k}{(1+k^2)^{3/2}}.
\end{equation}
This result is equivalent to the previous expressions if 
\begin{equation}
\int_0^1 \frac{{\mathbb{K}}(k)\,k\,{\text{d}}k}{(1+k^2)^{3/2}} = \frac{\pi}{4\!\sqrt2}.
\end{equation}
Neither Maple or Mathematica returns a symbolic evaluation of this integral, but it is easy to verify it numerically.

\subsection{The constant intensity sphere $m\rightarrow0$}
\label{ConstantIntensitySphere.sec}

In \citetalias{2019A&A...626A.110B}, we have discussed the structure of the family of S\'ersic models with a focus on the small S\'ersic indices, and also considers the special case of $m\rightarrow0$. This limiting model is characterised by a finite extent and a uniform surface brightness distribution,
\begin{equation}
I(R) =
\begin{cases}
\;I_0 &\qquad{\text{if }}R<\sqrt2\,\Re \\ \;0&\qquad{\text{if }}R>\sqrt2\,\Re.
\end{cases}
\end{equation}
This simple surface brightness distribution translates to a ball in which the density increases from the centre to an outer, infinite-density skin. Substituting the expressions (22) and (27) from \citetalias{2019A&A...626A.110B} for the density and mass profile into Eqn.~(\ref{Etot-def2}), we obtain the simple result 
\begin{equation}
E_{\text{tot}} = -\frac{8}{3\!\sqrt{2}\,\pi^2}\,\frac{GM^2}{\Re}.
\label{Etotm0}
\end{equation}
The same result can be found by taking the limit $m\rightarrow0$ in expression~(\ref{Etot-K}), if we take into account that $\lim_{m\rightarrow0} b^m = 1/\!\sqrt2$ \citepalias{2019A&A...626A.110B}. Since 
\begin{equation}
\lim_{m\rightarrow0}\, 
\frac{\Gamma(3m)}{m\,\Gamma^2(2m)}\,\frac{1}{(1+k^{1/m})^{3m}} = \frac43
\end{equation}
for all $0\leqslant k\leqslant1$, we find immediately that
\begin{equation}
E_{\text{tot}} = -\frac{8}{3\!\sqrt2\,\pi^2}\,\frac{GM^2}{\Re} \int_0^1 {\mathbb{K}}(k)\,k\,{\text{d}}k 
= -\frac{8}{3\!\sqrt2\,\pi^2}\,\frac{GM^2}{\Re},
\label{Etotm0}
\end{equation}
where the last transition follows from the fact that the integral is simply equal to unity \citep{Glasser1976}.

\section{Total energy for the most common models}
\label{AllModels.sec}

\begin{table*}
\centering
\caption{The total energy for the most commonly used spherical models. The upper half of the table, above the horizontal line, contains a number of popular one-parameter families of spherical models. The bottom half, below the horizontal line, lists a number of well-known specific models. The first column is the name of the model (or one-parameter family of models), the second column corresponds to either the density $\rho(r)$ or the surface brightness $I(R)$, depending on what is the most natural way to define the model. The third column is the total energy $E_{\text{tot}}$. Everything is expressed in normalised units, i.e., the gravitational constant, total mass, the mass-to-light ratio, and the scale length are set to one.}
\label{AllModels.tab}
\begin{tabular}{|c|c|c|}
\hline\hline & & \\[0.5em]
model & $\rho(r)$ or $I(R)$ & $E_{\text{tot}}$ 
\\[1.2em]
\hline 
& & \\
Dehnen or $\gamma$ & 
$\displaystyle{\rho(r) = \frac{3-\gamma}{4\pi}\,\frac{1}{r^\gamma\,(1+r)^{4-\gamma}}}$ &
$-\dfrac{1}{4\,(5-2\gamma)}$ 
\\[1.4em]
$\beta$ & 
$\displaystyle{\rho(r) = \frac{(\beta-2)\,(\beta-3)}{4\pi}\,\frac{1}{r\,(1+r)^{1-\beta}}}$ &
$-\dfrac{(\beta-3)^2}{4\,(2\beta-5)}$ 
\\[1.2em]
Veltmann &
$\displaystyle{\rho(r) = \frac{1+\lambda}{4\pi}\,\frac{r^{\lambda-2}}{(1+r^\lambda)^{2+1/\lambda}}}$ &
$\displaystyle{-\frac {\sqrt{\pi}\,(1+\lambda)} {2^{3+2/\lambda}\,\lambda^2}\,
\frac{\Gamma\left(\frac1\lambda\right)}{\Gamma\left(\frac32+\frac1\lambda\right)}}$ 
\\[1.2em]
Einasto &
$\displaystyle{\rho(r) = \frac{1}{4\pi\,n\,\Gamma(3n)}
\exp\left(-r^{1/n}\right)}$ &
$\displaystyle{-\frac{\Gamma(2n)}{2\,\Gamma(3n)}+\frac{\Gamma(5n)}{4n\,\Gamma^2(3n)}\,{}_2F_1(2n,5n;2n+1;-1)}$ 
\\[1.2em]
S\'ersic &
%$\displaystyle{\frac{b^{3m}}{\pi^{3/2}\,\Gamma(2m)}\,r^{-1}\,
%H^{2,0}_{1,2} \left[ \left.\begin{matrix} (0,1) \\ (0,2m), (\tfrac12,1) \end{matrix} \,\right| b^{2m}r^2 \right]}$ &
$\displaystyle{I(R) = \frac{b^{2m}}{2\pi\,m\,\Gamma(2m)}\exp\left(-bR^{1/m}\right)}$ &
$\displaystyle{-\frac{b^m}{2\pi\, \Gamma^2(2m)}\,H^{2,2}_{3,3}\left[\left.
\begin{matrix} (1-3m,2m),(0,1),(0,1) \\ (0,2m), (-\tfrac12,1), (-\tfrac12,1) \end{matrix}\,\right|1\right]}$ 
\\[1.5em] 
\hline 
& & \\
Plummer &
$\displaystyle{\rho(r) = \frac{3}{4\pi}\,\frac{1}{(1+r^2)^{5/2}}}$ &
$\displaystyle{-\frac{3\pi}{64}}$ 
\\[1.2em]
H\'enon's isochrone &
$\displaystyle{\rho(r) = \frac{1}{4\pi}\,\frac{1+2\sqrt{1+r^2}}{(1+r^2)^{3/2}\,(1+\sqrt{1+r^2})^2}}$ &
$\displaystyle{\frac13-\frac\pi8}$ 
\\[1.2em]
Hernquist &
$\displaystyle{\rho(r) = \frac{1}{2\pi}\,\frac{1}{r\,(1+r)^3}}$ &
$\displaystyle{-\frac{1}{12}}$ 
\\[1.2em]
Jaffe &
$\displaystyle{\rho(r) = \frac{1}{4\pi}\,\frac{1}{r^2\,(1+r)^2}}$ &
$\displaystyle{-\frac{1}{4}}$ 
\\[1.2em]
perfect sphere &
$\displaystyle{\rho(r) = \frac{1}{\pi^{2}}\,\frac{1}{(1+r^2)^2}}$ &
$\displaystyle{-\frac{1}{4\pi}}$ 
\\[1.2em]
constant density &
$\displaystyle{\rho(r) = \frac{3}{4\pi}\quad(r<1)}$ &
$\displaystyle{-\frac{3}{10}}$ 
\\[1.2em]
exponential &
$\displaystyle{I(R) = \frac{b^2}{2\pi}\,{\text{e}}^{-bR}}$ &
$\displaystyle{-\frac{3b}{32}}$ 
\\[1.2em]
gaussian &
$\displaystyle{I(R) = \frac{b}{\pi}\,{\text{e}}^{-bR^2}}$ &
$\displaystyle{-\frac{\sqrt{b}}{2\!\sqrt{2\pi}}}$ 
\\[1.2em]
constant intensity &
$\displaystyle{I(R) = \frac{1}{2\pi}\quad(R<\!\sqrt2)}$ &
$\displaystyle{-\frac{8}{3\!\sqrt2\,\pi^2}}$ 
\\[1.5em]
\hline\hline   
\end{tabular}
\end{table*}

In Table~{\ref{AllModels.tab}} we list the total energy for some of the most commonly used spherical models. The upper half of the table, above the horizontal line, contains a number of popular one-parameter families of spherical models. The bottom half, below the horizontal line, lists a number of well-known specific models. Each model is completely defined by either the spatial density profile or the surface brightness, and contains the total mass and a length scale (both of which have been set to one here) as free parameters.

\end{document}